\documentclass[pdflatex,sn-mathphys-num]{sn-jnl}

\usepackage{graphicx}%
\usepackage{multirow}%
\usepackage{amsmath,amssymb,amsfonts}%
\usepackage{amsthm}%
\usepackage{mathrsfs}%
\usepackage[title]{appendix}%
\usepackage{xcolor}%
\usepackage{textcomp}%
\usepackage{manyfoot}%
\usepackage{booktabs}%
\usepackage{algorithm}%
\usepackage{algorithmicx}%
\usepackage{algpseudocode}%
\usepackage{listings}%

\usepackage{subcaption}
\usepackage{adjustbox}
\usepackage{float}
\usepackage{array}

\theoremstyle{thmstyleone}%
\theoremstyle{thmstyletwo}%

\theoremstyle{thmstylethree}%

\begin{document}

\title{Combining Genre Classification and Harmonic-Percussive Features with Diffusion Models for Music-Video Generation}

\author{\fnm{Leonardo} \sur{Pina}}
\author{\fnm{Yongmin} \sur{Li}}

\affil{\orgdiv{Department of Computer Science}, \orgname{Brunel University London}, \country{UK}}


\abstract{This study presents a novel method for generating music visualisers using diffusion models, combining audio input with user-selected artwork. The process involves two main stages: image generation and video creation. First, music captioning and genre classification are performed, followed by the retrieval of artistic style descriptions. A diffusion model then generates images based on the user's input image and the derived artistic style descriptions. The video generation stage utilises the same diffusion model to interpolate frames, controlled by audio energy vectors derived from key musical features of harmonics and percussives. The method demonstrates promising results across various genres, and a new metric, Audio-Visual Synchrony (AVS), is introduced to quantitatively evaluate the synchronisation between visual and audio elements. Comparative analysis shows significantly higher AVS values for videos generated using the proposed method with audio energy vectors, compared to linear interpolation. This approach has potential applications in diverse fields, including independent music video creation, film production, live music events, and enhancing audio-visual experiences in public spaces.
Example results are available at \url{https://github.com/LeomPina/AI-Driven-Music-Visual-Generation-Using-Diffusion-Models.git}.}




\maketitle

\section{Introduction}

The pairing of visual and audio content is ubiquitous and continually evolving. In recent years, the use of artificial intelligence (AI) has been a game changer by helping to change the way we can perceive and experience these audio and visual representations. We are currently living in a digital era where music is not merely supposed to be associated with listening, since it is also related to a visual experience \cite{ref1} as with every major song release there is one correspondent music video. In fact, even in the early days of the music industry, the relationship between music and image was already present in the form of album artworks \cite{ref2}.

Audio and visual data have been integral to various fields, including multimedia, human-computer interaction, content analysis, multimodal processing, and visual perception, for many years \cite{xu2003video,li2001constructing,li2001video,li2000multi,li2003recognising,li2003constructing}.
Despite all the advancements that have occurred throughout the years, establishing a relationship between music and image is still very much a challenge as this association can easily become a subjective topic, in the sense that every person can adopt a different point of view and ultimately form a unique opinion about what visual representation best matches one particular piece of music. For this reason, there are still many unexplored approaches in the field of generating visuals from music using AI concepts and models. This underscores the importance of further research on this topic, as it could significantly influence how we visually perceive music.


Previously, studies have been conducted using text-to-image methods with diffusion models to produce images associated with a song sample \cite{ref3}, music-to-visual style transfer methods capable of using music to generate diverse image-based music representations going beyond the usual image style transfer problem \cite{ref4}, and the automatic image generation from music by learning the correlation between this data and using CNN, RNN, and GAN methods \cite{ref5}. Others have further explored how video could be generated through music, such as the use of lyrics, if existent, to perform image search and bridging a song with the images by calculating their matching semantic score and image quality to generate the music video \cite{ref6}. Another method involves automatically creating high quality music video by applying a pseudo song prediction method, while simultaneously suggesting soundtracks and editing user-generated videos (UGVs) \cite{ref7}. 

In this work, we propose a different solution for the music to visuals problem: our method allows users to choose a music sample along with a correspondent artwork image, as we acknowledge the association of a music piece with a visual representation can be an idiosyncratic and personal process. Instead of using pre-defined images for the creation of the visuals, our solution employs a diffusion model to generate different images that can capture the musical elements of the song (such as its genre and musical description) and, by using a large language model (LLM), translates this into data that will then heavily influence how these images are produced. Another diffusion model and algorithms are then used to take these output images to create the frames that will transition according to the musical components of the song, and finally complete the music video. This process is structured in two different stages, the music artwork image generation and the music visualiser generation.

This rest of the paper is organised as follows. In Section 2 we review related studies by providing an overview of existing literature. 
Section 3 discusses the details of our approach, divided into two main stages of image generation and music visualiser generation, where we explain the methods, AI models and algorithms used, and how they work together to form a coherent solution.
In Section 4 we present the results generated by our method, providing real examples of the direct application of our solution. We offer an evaluation analysis by using both qualitative and quantitative methods, while further discussing how our solution can address the underlying problem effectively and efficiently. We also highlight the obstacles and limitations encountered during different phases of the project.
Finally, conclusions are drawn in Section 5, where we offer an analysis on the entire process and an insight on what we were capable of achieving, along with some real application scenarios for this approach, key areas of future work and how these findings contribute to the improvement of this field of research.

\section{Background}

In this section, we provide a brief review of the previous work on the three main areas relevant to our approach to music-visual generation, from image generation, to video generation and music captioning. 

\subsection{Image Generation}
Diffusion models have been widely used for various computer vision tasks, surpassing the previously popular GANs on image sample quality \cite{ref8}, even though diffusion models require several forward passes during the sampling process as also addressed by \cite{ref9}. The latest target topic of the use of these diffusion models is described to be on the field of conditional and unconditional image generation. Text-to-image generation tasks have shown to be performed efficiently when using models such as Stable Diffusion and DALL-E2, models trained in latent space proposed by \cite{ref11} and \cite{ref12}, respectively, which achieved a good balance between image quality and text-image alignment evaluation criteria metrics \cite{ref10}. Diffusion models have commonly used CLIP \cite{ref27} to perform these text-to-image generation tasks, as CLIP models learn to be efficient in executing several types of tasks during pre-training and afterwards, are able to use natural language to create or refer to visual concepts, generalising its ability to unknown tasks. The CLIP-GEN \cite{ref13} model is also an example of the use of pre-trained CLIP models to perform text-to-image generation tasks, this time without the usage of paired text-image data, and an autoregressive transformer to create the images from the CLIP text embedding. Enhancements on the diffusion models, as the new state-of-the-art models for image generation, have been further pushed to adapt for many kinds of tasks. The Kandinsky model \cite{ref26} was developed to perform various tasks like text-to-image, image fusion, and text and image fusion generation by presenting a latent diffusion architecture. New diffusion image variation models have been proposed on recent studies, \cite{ref14} has proposed “Prompt-Free Diffusion” model to generate new output images based on example input ones, while still achieving high quality results. Comparably, the Stable Diffusion Image Variations Model \cite{ref33} also performs image variation generation tasks by employing stable diffusion, accepting CLIP image embeddings as input data (instead of text representations).

\subsection{Video Generation}
Apart from image generation tasks, diffusion models have been also implemented in the field of video generation as research work has been conducted in areas such as video generation, video editing, and video understanding \cite{ref15}. Proposed by \cite{ref16}, the state-of-the-art text-to-video generation diffusion model based itself on the standard image diffusion structure by using U-Net architectures allowing a simultaneous training on both image and video data. This model employed a new conditional sampling technique to produce high quality and long videos by using frame interpolation or extrapolation, further making it possible to execute spatial super-resolution. Imagen Video \cite{ref17} was presented as a text-conditional video generation system that accepts text prompts as input data to generate high definition videos basing itself on a sequence of several diffusion models, while also displaying a high level of controllability, world knowledge, and temporal consistency, which was achieved with the use of fully-convolutional temporal and spatial super-resolution models.

The use of audio to generate image and video is still considered a challenging area, nevertheless, many studies have been lately conducted to create adequate solutions. Earlier studies have presented a solution, using user’s personal home videos that analysed and matched the visual structure and patterns of the video with those same aural components existent in the song to finally generate a music video with the appropriate transition effects \cite{ref19}. To study how the relation between music and image could be transformed into visual content, \cite{ref6} created a system that is able to produce a music video by using a certain song, employing methods such as internet image search, music-image matching, and photo quality evaluation. The described approach also presents user controllable features as the user can choose his own images. More recently, the text-to-video AADiff \cite{ref18} framework showed how audio-aligned videos can be synthesised with the use of text and audio signals (for example, a music sample) as inputs, a text-to-image diffusion model, an audio-based regional editing and signal smoothing between frames technique to achieve a balance between the video’s temporal flexibility and coherence. Concomitantly, \cite{ref3} proposed a music visualisation pipeline method by also using a text-to-image diffusion model to generate the final image representations. This pipeline accepts a music sample and user chosen artwork images, as both fed data are transformed into text by utilising music-to-text and image-to-text models to be used as input data for the diffusion model. The process of successfully transforming a music sample into a text form that coherently represents its essence, as it can then be used to create other forms of content, is still a challenge.

\subsection{Music Captioning and Tagging}
\hspace*{\parindent}Existent large language models have also been built to transform music samples into text representations. Accordingly, the musicnn \cite{ref20} library is a method built to perform music-to-text tasks, more specifically, music audio tagging generation as it is capable of predicting the music tags of a song. To successfully achieve this, this library is composed of pre-trained convolutional neural networks, while also presenting vgg-based models. Similarly, many other methods have been developed to efficiently execute these music tagging and captioning tasks, for instance, the first audio captioning model employed a network composed of a multimodal CNN-LSTM encoder and an LSTM decoder to generate text descriptions that capture the content of a music sample \cite{ref21}. Based on the CLIP model, the AudioCLIP \cite{ref22} model extended CLIP to also handle audio representations by using a new ESResNeXt audio model for audio classification. More recently, the LP-MusicCaps \cite{ref23} large language model-based was developed to perform automatic music captioning tasks, while using large-scale tag datasets to effectively generate pseudo captions associated to a given music sample. This detailed description generation approach consists in the creation of the LP-MusicCaps dataset that was then also used to train a transformer-based music captioning model, which showed better results than the existent literature related to this field of research. To further explore and address the gap between the text and music relation domain, the MusiLingo \cite{ref24} system was created to perform music caption and music related question-answering tasks by using the music representations obtained from a pre-trained frozen music audio MERT model in a single projection layer alongside a frozen large language model, while introducing a new music caption dataset to improve its efficiency.

\subsection{The Proposed Work}
Diffusion models have been widely used for generating images and videos from various input of text, image, and audio. These models, along with large language models, have helped link music and text. However, current methods still face limitations and challenges such as the lack of user guidance, personalisation, the music's constant changes, and inefficient evaluation techniques. To address these problems, we introduce a two-stage method for generating music videos from a music input and a user chosen artwork image. The key contributions of the work are as follow:
\begin{enumerate}
\item 
We introduce an integrated approach to image generation through music captioning, genre classification, and the application of a diffusion model to create images based on user-selected artwork and the identified artistic style.
\item 
The audio energy vectors, derived from key musical features such as harmonics and percussives, are introduced to control the interpolation of frames during video generation. This innovative approach enhances the visual representation's alignment with the music.
\item 
A new metric, Audio-Visual Synchrony (AVS), is developed to quantitatively assess the synchronisation between visual and audio components in generated music videos. 
\end{enumerate}


\section{Methods}

\begin{figure}[H]
    \centering
    \includegraphics[width=0.9\linewidth]{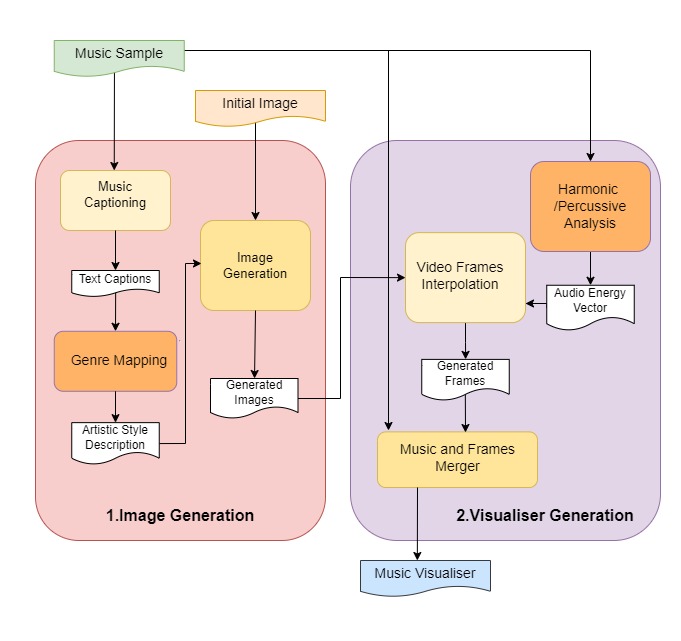}
    \caption{The overall block diagram of the proposed approach,  with two main stages of image generation and video synthesis.}
    \label{fig:project_arch}
\end{figure}

As shown in Figure \ref{fig:project_arch}, the proposed approach involves two main stages of image generation and video synthesis. 
The input to the system is a music piece and an initial seed image chosen by the users. First, the music piece is fed into a music captioning module to extract text description of the music, then the captions are classified into one of the pre-defined music genres, and finally a genre-wise artistic style description of the music is retrieved from the genre dictionary using the genre as the key. In the final step of the image generation, the artistic style description and the seed image are combined as input to a diffusion model to generate images, one for each music segment, e.g. 10 seconds.

In the second stage of video synthesis, the harmonic and percussive components of the music piece are estimated to form an audio energy vector, which later controls the individual frame generation of the music video through another diffusion model, and these frames are merged into the final music video.


\subsection{Music Captioning}
For the first stage of image generation, the objective is to use one music sample chosen by the user and one or more associated artwork images as inputs to generate new AI images that would be capable of displaying and ultimately combining the musical content of the audio sample with the initial image(s). 


We adapted the LP-MusicCaps \cite{ref23} for initial music captioning, i.e. to generate a text caption for each segments of the music piece. The length of segments is set to 10 seconds in our experiments. 
Built on the the GPT-3.5 Turbo LLM model presented by \cite{ref25}, the LP-MusicCaps used a cross-modal encoder-decoder transformer architecture, where the encoder accepts a 10 seconds Log-mel spectrogram. The encoder is composed of six convolutional layers and uses a GELU activation function. The output from these convolutional layers is then merged with a sinusoidal position encoding method (ensuring unique encodings). The 6 transformer blocks that compose the encoder then process the previous data. While following a similar BART architecture, this model’s decoder will take the tokenised text captions and process them through the transformer blocks, while incorporating a multi-head attention module that applies a mask to conceal future tokens to ensure causal prediction. Finally, the cross-modal attention layer will process the representations obtained from the audio sample and the text captions, by using the language model’s head, which  then predicts the next text token autoregressively.

By using this model, we were able to successfully generate text captions that describe the overall summary and different elements presented in the music sample. The example in Table \ref{tab:30sec_song_caption} shows obtained results when applying this model to one electronic music.

\begin{table}[]
    \centering
    \begin{adjustbox}{width=\textwidth}
    \begin{tabular}{|c|p{7cm}|}
    \hline
    Time (seconds) & \multicolumn{1}{c|}{Description} \\
    \hline
    0:00-10:00 & The song is an instrumental. The tempo is medium with a strong bass line, steady drumming rhythm, triangle percussion, keyboard accompaniment, and other percussion hits. The song is a modern techno dance tune. \\
    \hline
    10:00-20:00 & This song contains a digital techno drum with a kick on every beat. A sub-bass is inviting to dance along with synth pluck sounds playing a melody in a higher key along with a female voice. A very low-pitched male voice sample comes in for a moment before a synth pad sound playing in the midrange comes in. This song may be playing in a techno club. \\
    \hline
    20:00-30:00 & This music is an electronic instrumental. The tempo is fast with synthesizer arrangements, electronic beats, DJ mixer, and a repetitive vocal riff. The music is incessant, psychedelic, hypnotic, trippy, and trance-like with a dance groove. This music is Techno Pop/EDM. \\
    \hline
    \end{tabular}
    \end{adjustbox}
    \caption{Examples of the music captioning results from  30-second sample of an Electronic music.}
    \label{tab:30sec_song_caption}
\end{table}

\subsection{Music Genres to Guide the Image Generation}
To achieve more controllable and consistent results when generating images, we opted to use a set of pre-defined music genres to guide how the model generates the output images related to the chosen song sample and associated artwork.

We first identified a total of 23 labels of the more commonly known genres such as “pop”, “rock”, “punk”, “jazz”, “blues”, “indie”, “classical”, “country”, “folk”, “electronic”, “techno”, “electro”, “house”, “gospel”, “latin”, “metal”, “rap”, “hip hop”, “reggae”, “reggaeton”, “funk”, “disco” and “R\&B”. 
Then we created a description based on different art styles, moods and other visual elements that could be easily associated to each genre. To get clearer results, some sub-genres were considered as the same and therefore were assigned the same description. Table \ref{tab:music_genres} shows the text descriptions that were assigned to each of the genres considered.

\begin{table}[]
    \centering
    \begin{adjustbox}{width=\textwidth}
    \begin{tabular}{|l|p{12cm}|}
    \hline
    Music Genre & Description \\
    \hline
    Pop     &  Pop Art with bright, bold colors, high contrast, and a playful, energetic vibe. Think of the works of Andy Warhol and Roy Lichtenstein\\
    \hline
    Rock     &  Grunge and Expressionism with chaotic, intense, and emotive elements. Darker color palettes with splashes of red and black\\
    \hline
    Punk     &  Grunge and Expressionism with chaotic, intense, and emotive elements. Darker color palettes with splashes of red and black\\
    \hline
    Jazz     &  Abstract Art, often with flowing lines and dynamic compositions. Use of vibrant yet deep colors, echoing the spontaneity and complexity of jazz\\
    \hline
    Blues     &  Realism with a focus on somber and moody tones. Use of sepia tones and muted colors to evoke a sense of nostalgia and depth\\
    \hline
    Indie    &  Indie Art and Folk Art with whimsical, eclectic, and often minimalistic elements. Use of pastel colors and hand-drawn illustrations\\
    \hline
    Classical     &  Impressionism with delicate brushstrokes and soft color palettes. Think of the works of Claude Monet and Pierre-Auguste Renoir\\
    \hline
    Country     &  Americana and Western Art with earthy tones and rural themes. Think of wide-open landscapes and vintage illustrations\\
    \hline
    Folk     &  Folk Art with simple, rustic, and often narrative elements. Use of earthy colors and traditional patterns\\
    \hline
    Electronic     &  Futurism and Digital Art with neon colors, geometric patterns, and a high-tech aesthetic\\
    \hline
    Techno     &  Futurism and Digital Art with neon colors, geometric patterns, and a high-tech aesthetic\\
    \hline
    Electro     &  Futurism and Digital Art with neon colors, geometric patterns, and a high-tech aesthetic\\
    \hline
    House     &  Futurism and Digital Art with neon colors, geometric patterns, and a high-tech aesthetic\\
    \hline
    Gospel     &  Renaissance and Religious Art with heavenly light, vibrant colors, and ethereal themes. Think of stained glass and divine imagery\\
    \hline
    Latin     &  Muralism and Latin American Folk Art with bold colors, dynamic compositions, and cultural motifs\\
    \hline
    Metal     &  Gothic and Heavy Metal Art with dark color palettes, skulls, flames, and powerful, dramatic imagery\\
    \hline
    Rap     &  Street Art and Graffiti with bold colors, sharp lines, and urban themes. Think of murals and hip-hop culture\\
    \hline
    Hip Hop     &  Graffiti and Pop Art with bright colors, dynamic compositions, and cultural references. Think of vibrant murals and street culture\\
    \hline
    Reggae   &  Rastafarian Art with warm colors, tropical themes, and cultural motifs. Use of greens, yellows, and reds\\
    \hline
    Reggaeton     &  Urban Art with bright colors, dynamic patterns, and Latin influences. Think of vibrant cityscapes and dance culture\\
    \hline
    Funk     &  Psychedelic Art with vibrant colors, dynamic patterns, and funky, retro elements. Think of the works of Peter Max\\
    \hline
    Disco     &  Psychedelic Art with vibrant colors, dynamic patterns, and funky, retro elements. Think of the works of Peter Max\\
    \hline
    R\&B     & Neo-Soul Art with warm colors, elegant lines, and romantic themes. Use of deep blues, purples, and golds \\
    \hline
    \end{tabular}
    \end{adjustbox}
    \caption{Music genres descriptions.}
    \label{tab:music_genres}
\end{table}

The text descriptions shown above were then put in the form of a dictionary that represented the correspondence between the genre labels (keys) with its descriptions (values). 
We then performed a simple text classification to determine the music caption to a specific genre, and used the genre as the key to extract its corresponding description as the artistic style description for the next step of image generation.


For the image generation, we used a diffusion model based on the Kandinsky model \cite{ref26} that takes both the artistic style description and the initial artwork images as input. 
The Kandinsky model presents an architecture based on latent diffusion, while adopting some image prior models’ concepts. This model is composed of three main sections, text encoding, embedding mapping, and latent diffusion. To obtain a better performance and higher image quality, both CLIP-text with image prior mapping and XLMR are used for text encoding. For the embedding mapping (image prior), the model uses a transformer-encoder model, which were trained with image and text embeddings from the CLIP models \cite{ref27}. A UNet model alongside a pre-trained autoencoder are used in the latent diffusion section, while also combining different components such as CLIP-image embeddings (input), CLIP-text embeddings, and XLMR-CLIP text embeddings (input). This model’s Sber-MoVQGAN autoencoder is an alternative to MoVQGAN developed by \cite{ref28}. 

Figure \ref{fig:Diff_music_genres_same_image_results} shows how the genre information would affect the generation of different images when paired with the same initial input image,
where generated images are from the first 30 seconds with an interval of 10 seconds.
It can be observed that when opting for different genres like “classical”, “jazz” or “rock”, we get an image more relatable to its sample’s musical content, either more aggressive and dark elements for the “rock” genre, smooth and delicate for “classical”, or abstract and complex for “jazz”. We found this approach to be more visually appealing than simply using the text summary from music captioning as input. Video examples that use this technique’s resulting images can be found in section \ref{sec:Experiments and Results Analysis}.

\subsection{Audio Energy Vectors}
For the second stage, the aim was to take the images generated in the previous stage and use them to create the video animation. In order to make this possible, we used a stable diffusion image variations model \cite{ref33} to transform the images from one to another and therefore create movement and animation. To guide the flow of this movement and transitions, we used the librosa python package and the ffmpeg python library to put everything together, the audio sample with the generated frames. This stage of our research was based on the work done by \cite{ref32}.

\begin{figure}
    \centering
    \begin{subfigure}[t]{0.48\textwidth}
        \centering
        \includegraphics[width=\textwidth]{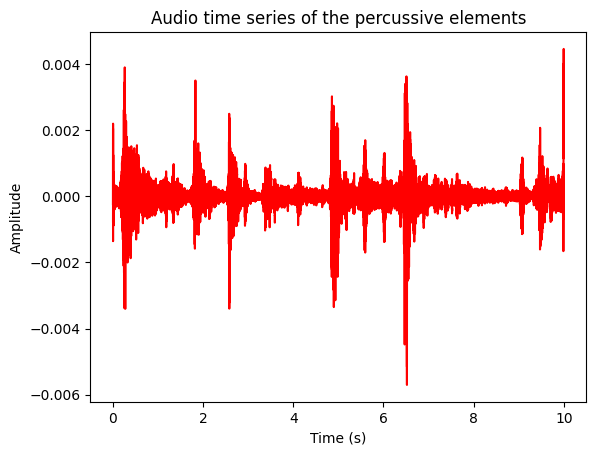}
        \caption{}
    \end{subfigure}
    \begin{subfigure}[t]{0.48\textwidth}
        \centering
        \includegraphics[width=\textwidth]{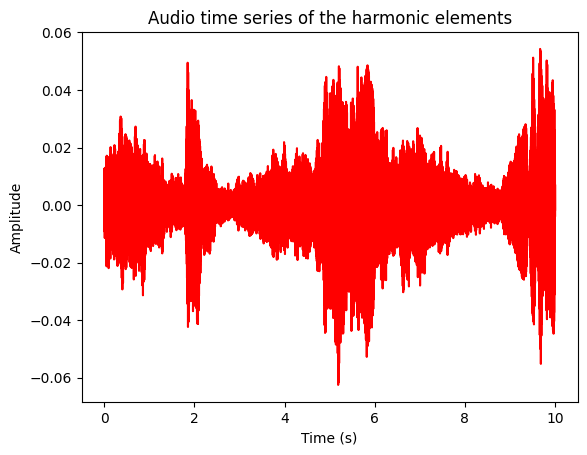}
        \caption{}
    \end{subfigure}
    \begin{subfigure}[t]{0.48\textwidth}
        \centering
        \includegraphics[width=\textwidth]{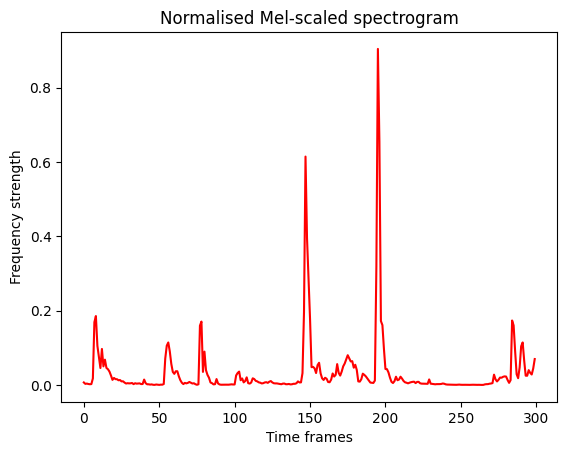}
        \caption{}
    \end{subfigure}
    \begin{subfigure}[t]{0.48\textwidth}
        \centering
        \includegraphics[width=\textwidth]{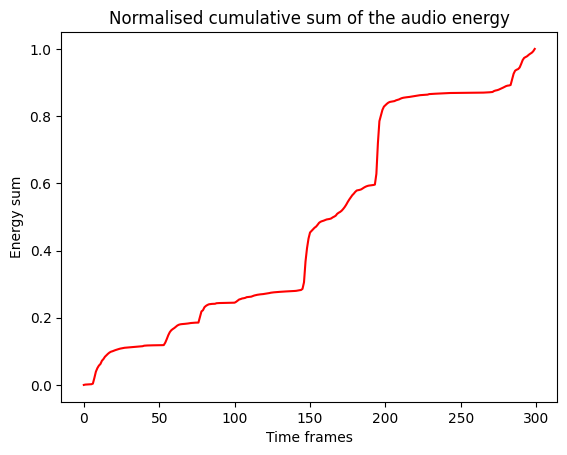}
        \caption{}
    \end{subfigure}

    \caption{The audio energy vector used to control the synthesis of a 10-second music clip. (a) percussive and (b) harmonic audio time series, (c) normalised Mel-scaled spectrogram and (d) the final audio energy vector.}
    \label{fig:perc_harmon_clips}
\end{figure}

So far we have a generated image for each segment, e.g. 10 seconds, of the music. In the following steps, all frames of the output video are interpolated in the latent variable space and then generated by a diffusion model \cite{ref33,ref32}. Instead of using a linear transform for the interpolation, we introduce an Audio Energy Vector to control the flow of image transition. 
The basic idea is to align the image changes in the video to the acoustic 
features such as beat, tempo or even the overall musical elements like harmony or melody. First, we decompose an audio time series into harmonic and percussive components 
as shown in Figure \ref{fig:perc_harmon_clips} (a) and (b) respectively.

The percussive component displays more intense and accentuated amplitude transitions, showing more peaks, while the harmonic is more balanced and smooth. 
These two components are combined together with different weights 
to calculate the Mel-scaled spectrogram and cumulative sum of the audio energy \cite{xu2003video,xu2005method,li2001dynamic}. 
Figure \ref{fig:perc_harmon_clips} (c) and (d) show the Mel-scaled spectrogram and the audio energy vector using weights of 0.9 and 0.1 for percussive and harmonic respectively.

\begin{figure}[H]
    \centering
    \includegraphics[width=\linewidth]{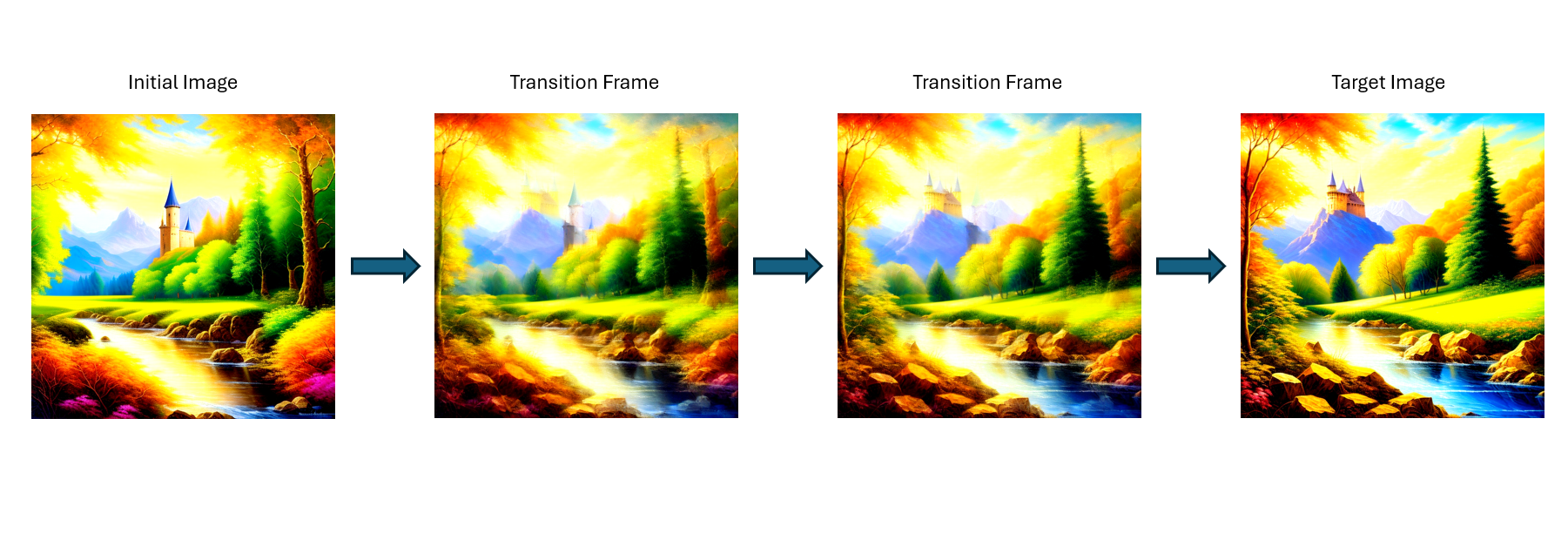}
    \caption{Transitional frames generated from an initial into a target image.}
    \label{fig:transitional_frames}
\end{figure}

The cumulative audio energy vector is then used as a guide to smoothly interpolate between two images (latent vectors), resulting in the intermediate frames. In other words, value 0 in the cumulative sum represents the first frame, 1 the final frame, and others the frame transitions generated from the interpolation of the initial image to the final image. This process is repeated until all frames of the video are produced. Figure \ref{fig:transitional_frames} shows how this transition can visually happen during the 10-second duration period with the help of the model presented in the next section.

\subsection{Frame Interpolation}
We use the same diffusion model as in the previous image generation step, except for no need of input text.
The latent vectors of all the intermediate frames generated with this diffusion model are created with a slerp (spherical linear interpolation) function and then being decoded back into an actual image using the VAE (variational auto-encoder) decode pipeline (the encoding of the initial and target images had as well previously occurred with the use of the VAE encode pipeline).


\subsection{Visualiser synthesis}

To finally create the music visualiser, all the previously interpolated and generated frames are mixed together with the music input into a final video. This can be implemented by any video encoding software, such as the ffmpeg library \cite{ref41}.

\section{Experiments and Results Analysis}
\label{sec:Experiments and Results Analysis}
In this section, we will detail the experiments conducted and provide a thorough evaluation of the results obtained. Each of the produced visualisers has a frame rate of 30 frames/second and a duration of 30 seconds. Samples of the results are available at
\url{https://github.com/LeomPina/AI-Driven-Music-Visual-Generation-Using-Diffusion-Models.git}.

\begin{figure*}
    \centering
    \includegraphics[width=\linewidth]{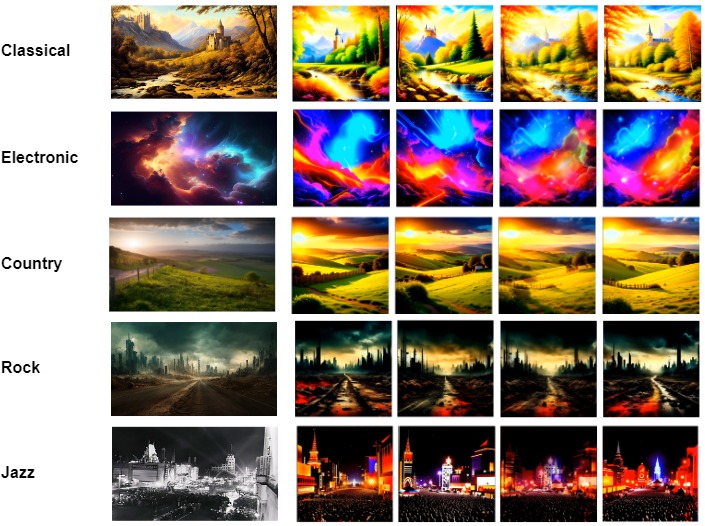}
    \caption{Generated results for each music genre with a related initial artwork image. From left to right are the music genre, initial image and generated images for each 10 seconds.}
    \label{fig:music_genres_results}
\end{figure*}

Figure \ref{fig:music_genres_results} shows the generated music video example results that we obtained for each  music genre, displayed along with initial artwork images.

\begin{figure}
    \centering
    \includegraphics[width=0.6\linewidth]{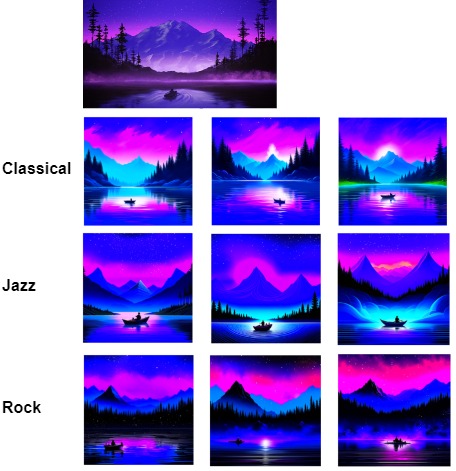}
    \caption{Generated results using the same initial image (top) for different genres.}
    \label{fig:Diff_music_genres_same_image_results}
\end{figure}

The purpose of this last result example is to further show how different music genre samples will affect the same initial artwork and consequently lead to the generation of several different visualisers. 
For this scenario we used the resulting images generated for different genres as input images to build the music animations, it’s important to notice that all these images were generated by using the same initial artwork image shown in Figure \ref{fig:Diff_music_genres_same_image_results} as input.

Figure \ref{fig:Diff_music_genres_same_image_results} shows the music video example experiments that we generated for different genres using a same initial artwork image.

To evaluate the obtained results shown before, we performed two different types of data evaluation, the qualitative and the quantitative evaluation, as the first one is based more on a direct observation and the second requires a real numerical metric.

\subsection{Qualitative Evaluation}
\hspace*{\parindent}We first employed a qualitative evaluation to our generated results that comprises a subjective critical analysis of these outputted results. The initial artwork image chosen by the user is ideally meant to be a heavy influence in the subsequent created animation, which means that the frames presented in the video should look related to the initial artwork. By looking at our experimental results, we can say that the images seen throughout the videos do look like they were based in the initial one when it comes to colours and the overall scenario displayed in this first picture. For example, when looking at Figure \ref{fig:music_genres_results}, for the classical song experiment, we can notice that the frame images share very strong similarities with the correspondent initial image, since we can still see in these frames the main elements that compose the artwork one, such as its bright colours, the mountain, the river and the castle. The same effect is also visible in the other video frames that were based in the initial images displayed in Figure \ref{fig:music_genres_results}.

Despite the new elements added to the AI generated images, the lack of controllability is still an issue when it comes to trying to guide the generation process, as the new images might not align to the content the user would expect or hope for when running the model, causing some dissatisfaction on the content produced. Due to the fact that the music genres also have a big influence in the images rendered, the user should choose an artwork that has something to do with the music’s genre so this algorithm’s feature can be a more useful tool, instead of leading to less clear, consistent and visually appealing images. By looking at our experiments in Figure \ref{fig:music_genres_results}, we can see that for different music genres, the user can choose an image artwork that can better relate to it, as for a classical song we chose a peaceful oneiric scenario, for an electronic song we chose a more futuristic and space related artwork, for a country song we picked a rural and natural field scenario, for a rock sample we selected a more aggressive and apocalyptic road scenario, and for a jazz song we picked a full of life vintage city landscape. The results presented by Figure \ref{fig:Diff_music_genres_same_image_results} show that for the same lake scenario image artwork, the use of a classical piece produces a more soft and bright image, a jazz song creates more arbitrary and confusing shapes, meanwhile, a rock song generates darker and more aggressive tonalities, which are coherent visual effects according to the prompts shown in Table \ref{tab:music_genres}. On the other hand, this feature can also be a problem as it could wrongly identify the music’s genre or offer an inaccurate text description. We believe that the visualisers built correspondent to the classical, electronic, country, rock, and jazz genres (displayed in Figure  \ref{fig:music_genres_results}) show a coherent transition flow between frames and the interpolation visualised between the inputted images, the user can adjust the fps parameter to make this transitions smoother or rougher, depending on the desired result. The music used alongside these video animations blends well with the frame transitions, as we have confidence that the viewer can easily feel the relation between the musical elements in the audio sample and the visuals displayed. Moreover, we are convinced that the results previously presented are able to show clearly this dynamic that whenever the song registers stronger frequencies (of a certain component, could be either harmonic or percussive), this intensity is also seen in the way the visualiser images are changing, so either more abruptly or smoothly.

\subsection{Quantitative Evaluation}
\begin{figure}
    \centering
    \begin{subfigure}[t]{0.48\textwidth}
        \includegraphics[width=0.95\linewidth]{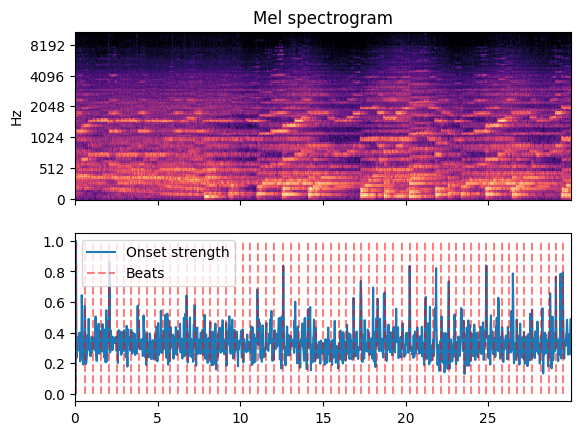}
        \caption{}
    \end{subfigure}
    \begin{subfigure}[t]{0.48\textwidth}
        \includegraphics[width=0.95\linewidth]{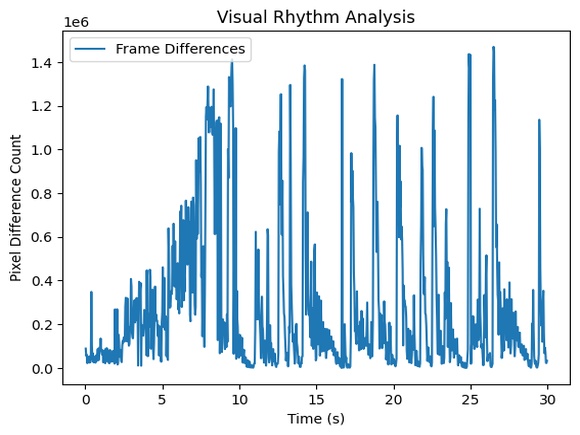}
        \caption{}
    \end{subfigure}
    \caption{Audio and visual dynamics of Classical song. (a) Mel spectrogram, (b) Pixel frame differences.}
    \label{fig:mel_spectogram_classic}
\end{figure}

As we do not have any ground-truth video examples or other comparison method,  we developed our own metric to evaluate the results quantitatively. For this purpose, we created the Audio-Visual Synchrony (AVS) metric, designed to measure how well the transitions in the visual content are synchronised to musical changes.
In order to do so, we first detect the significant beats existent in the audio sample by extracting the onset strength with the help of a threshold value (it can be edited) and then store the timestamps (in seconds) correspondent to when this happened in the sample. The plot presented in Figure \ref{fig:mel_spectogram_classic} (a) shows an example of how a music piece can be visually represented in terms of the intensity of its beats, the x-axis contains the duration of the song in seconds.

Meanwhile, to analyse the video and detect abrupt visual changes, we created an algorithm capable of performing this analysis by calculating the differences between consecutive frames, saving the correspondent timestamp (in seconds) if the calculated value reveals itself to be higher than the threshold value (it can be edited). Whenever these timestamp values are stored, the value of the slope correspondent to the linear regression, obtained by the previous calculated differences between consecutive frames related to the section in between these recorded timestamps, the algorithm is also responsible for storing these values. Another parameter is also used to prevent too many transition detections to take place in a short period of time, which could make this analysis less efficient, both these parameters make it possible to only consider and store major transitions. Figure \ref{fig:mel_spectogram_classic} (b) presents how this video frame transitions can be visually displayed. 


After performing all the mentioned analysis, we apply the Dynamic Time Warping (DTW) to compare the degree of similarity when comparing both the extracted timestamp vectors from each method (audio and video). The DTW method \cite{ref34} is an algorithm created to be capable of calculating the distance between two distinct time series by applying the dynamic programming technique to find the optimal temporal matching between the values of these two time series. Used widely in fields such as speech recognition \cite{ref35}, this algorithm calculates the similarity or dissimilarity of two time series by transforming the data into vectors and then measuring the absolute distance, which is normally the Euclidean distance, between those points. The Euclidian distance can be expressed mathematically in the following expression:

\begin{equation}
    d(p,q)=\sqrt{(p-q)^2},
\end{equation}
where p and q both represent the point values and d the distance between them (in a one dimension plane). Assuming that we have two numerical sequences such as $X=x_1,x_2,\dots,x_i,\dots,x_n$ and $Y=y_1,y_2,\dots,y_i,\dots,y_m$, this means that we can obtain two sequences with different lengths. The optimal warping path is the mapping of the values in both sequences to minimise the distance between them, which can also be described as a sequence of grid points $(i,j)$. This minimal distance matrix (cost matrix) between sequences can be computed by using the following recursive expression:
\begin{equation}
    D(i,j)=d(x_i,y_i)+min(D(i-1,j-1), D(i-1,j), D(i,j-1))
\end{equation}
where $d(x_i,y_j)$ is the Euclidian distance between $x_i$ and $y_j$, and $D(i,j)$ represents the cost of aligning (minimal distance) the subsequences $x_1,\dots,x_i$ and $y_1,\dots,y_j$. The optimal warping path cost of the two sequences can be expressed through the mathematical expression shown below:
\begin{equation}
    TC=\sum_{(i,j)\in Optimal Path}d(x_i,y_j)
\end{equation}
Or simply,
\begin{equation}
    DTW(X,Y)=D(n,m)
\end{equation}
where TC is the total cost associated to that path, X and Y are two sequences, and n and m are the lengths of these sequences, respectively. This will finally tell us what the total DTW cost is. In our project we implemented an approximate DTW algorithm optimised for faster computation, the FastDTW (Fast Dynamic Time Warping) \cite{ref36} that offers optimal (or near optimal) alignments, while having a near-linear time complexity. This alternative DTW algorithm can further show enhancements by being more memory efficient or even handling larger data.

This evaluation method would be more flexible and efficient towards rating video animation examples that do not present enough movement and frame transition animation, we implemented a weight penalty for these cases. Once the DTW value got calculated, the algorithm uses the previously slope stored values and calculates the mean associated to all theses values, so we can have one clear definitive value to work with while calculating the penalty weight. To calculate this penalty, we used an exponential function with an associated scaling factor, which normalises the final value, in a scale of 0 to 1, as the final rhythm consistency value obtained before by the calculation of the DTW value is then multiplied by this penalty, leading the algorithm to produce a final score. It is to be expected that results with a poorer animation in terms of lack of movement or too much predictability, would present lower penalty weights (closer to 0).


Table \ref{tab:audio-visual_results} shows the comparison of the AVS values of a selection of generated music videos of different genres using our methods against the simple linear transition. 

\begin{table}
    \centering
    \begin{tabular}{|p{2cm}|>{\centering\arraybackslash}p{2.6cm}|>{\centering\arraybackslash}p{2.6cm}|}
    \hline
    Genre     &  Proposed method & Linear interpolation\\
    \hline
    Classical     &  0.744 & 0.429\\
    \hline
    Electronic      &  0.738 & 0.501\\
    \hline
    Country     &  0.631 & 0.510\\
    \hline
    Rock      &  0.645 & 0.373\\
    \hline
    Jazz      &  0.748 & 0.311\\
    \hline
    \end{tabular}
    \caption{Audio-Visual Synchrony (AVS) evaluation results, with comparison between videos synthesised by the proposed method using the audio energy vectors and those without by linear interpolation.}
    \label{tab:audio-visual_results}
\end{table}

An analysis of the results in Table \ref{tab:audio-visual_results} reveals promising outcomes.  Acceptable results as obtained with the AVS values close to 1, showing that the video animations are well synchronised with the musical content heard in the music samples. Furthermore, it is noticed the AVS values are significantly lower with linear interpolations, indicating that those visualisers are not adequately synchronised with the musical changes.


\subsection{Challenges and Limitations}
\label{subsec:Challenges and Limitations}


The proposed approach, whilst successful in generating semantically synchronised music videos, is not without its limitations. The image generation using diffusion models along with video generation are recent fields of research, so the models available are still very limited and constantly updating. At the same time, the use of music to generate images and video animations is also an uncharted topic, as even today researches are not entirely sure how to relate a music piece to a certain image or scenery. For this reason, our first issue when starting to conduct our studies was not knowing how to establish the connection between a song and an image, or how the final visualiser would express the music’s content. The models used in this project were not always very efficient, the text descriptions generated by the LP-MusicCaps model throughout the music sample were at times inaccurate, as the genres mentioned and or the musical content summarised was not correct. This inefficiency was an issue that would heavily influence the image creation process using the Kandinsky model, leading to possibly unwanted resulting images. To try to tackle this issue and minimise the effect the text inputs had in the diffusion model, we lowered this input’s weight to only 0.30 to produce our proposed examples, forcing the model to prioritise the initial artwork image input that had a weight of 0.70 (this for all these results), which we found to be quite useful when dealing with the lack of control when it comes to this image generation procedure. Nevertheless, this insufficiency of control when using the Kandinsky model was a complication, as we feel that the user should always be able to guide the process towards a desirable result. 
For this reason, we created the genre mapping, so we could expect certain elements and design styles to be present in the outputted new image. However, our mapping currently only included the most commonly known musical genres. Consequently  if the music captioning model identifies a genre that is not included, the method will be unable to detect it and associate it to certain visual features. 

When developing the second stage of this project, some obstacles were also encountered, more precisely when creating the music videos. By using the librosa package to extract the sample’s musical elements, we can only do so for the harmonic and percussive components (these are transformed into audio time series), which limited our options when it comes to choosing which musical elements would influence and guide the most the flow of the frame transitions in the outputted visualiser. The Stable Diffusion Image Variations Model used to create the frames presented limitations such as the quality of the images generated could never be considered photorealistic (a limitation also present in the Kandinsky image generation diffusion model), both diffusion models used in this project for the task of image generation don’t perform well when dealing with pictures that contain human content as the model will usually output images that display, for example, limb or finger malformation, deformed faces, among other ugly and incoherent features. The model used in this second stage presents a lossy autoencoding process and reveals itself to be social biased since it was primarily trained on images associated to English descriptions, which can lead to a certain negligence towards texts and images from other cultures. When creating the video frames, the algorithm needs one initial input image and since our method only produces a new image for each 10 seconds, the user needs to pick himself this initial image to start the process, which makes it more toilsome. Music pieces that presented stronger percussive elements (such as more obvious or accentuated drum patterns and simple beats) showed clearer transition patterns that could be more easily seen in the final music video, for example, we noticed that the jazz genre, when compared to others, led to more unpredictable image transitions in the outputted results. This means that the song choice can heavily influence the quality of the visuals created.

In order to be able to evaluate this project’s results, we found it challenging to find an efficient evaluation method that could measure how acceptable our final visualiser results were. We decided to implement a metric that would actually assess how consistently the musical segment aligned with the visual part instead of just considering how aesthetically pleasing the video looked or its overall image quality, since this type of evaluation can be easily considered subjective. Even though the developed Audio-Visual Synchrony evaluation method used to assess our results was insightful, clear and consistent, it still presented some weaknesses. Both the functions to detect the significant beats existent in the audio sample and the major frame transitions in the visualiser use a threshold parameter value that is used so the algorithm only registers the timestamps of the values above that one. By doing so, there are still a lot of values that could be important for the analysis but their timestamps are not stored, leading to poorer evaluation results. When developing this metric, trying to synchronise the thresholds between these two functions so that these meaningful values recorded in both domains (visual and audible) were correspondent to each other was a great obstacle. The music choice can influence the metric results depending on the musical content that composes the sample as stronger elements tend to get extracted with ease. The DTW method applied in this evaluation process when comparing the similarity between two different time series can be sensitive to noise and outliers existent in the data causing inaccurate results. This method focuses on measuring global similarities between two sequences possibly ignoring local similarities as this algorithm has the objective to try to make an association between each part of one sequence and another part present in the other one, in our specific scenario, always considering the whole music piece and animation instead of just some key sections that correlate better with each other.

\section{Conclusions}


In this work, we have developed a new method to generating visualisers from a combined input of a music piece and an artwork image of the user's choice. 

In the first stage of image generation, music captioning is first performed on the input music, followed by a text classification into one of the pre-defined music genres, and then the artistic style description is retrieved from the genre dictionary. A diffusion model is developed to generate images from both the input of the user's initial image and the artistic style description obtained from the previous step for each small segments of the music.

In the next stage of video generation, we use the same diffusion model to interpolate all the middle frames between the first and last images of the music segment. This interpolation process is controlled by the audio energy vectors we introduced, which are formed from the key musical features of harmonics and percussives. 

We have presented promising results with different genres with the proposed method. In order to quantitatively evaluate the results, we have also introduced a new metric AVS to measure the synchrony between the visual and audio parts of the generated music video. Comparison results indicated that the generated music videos by our method with the audio energy vectors achieved significantly higher AVS values than those without, e.g. the linear interpolation.

The proposed approach has significant real-world application across various domains: to help independent artists creating their own music videos for streaming or social media platforms; to assist in the production of film or music video content; to enhance the live music user experience as it can be used as background visuals that correspond to the music listened in a music festival concert, or other live events; to improve the music-visual experiences in spaces such as fitness clubs, museums, and other public places, which can help in the engagement and immersion of the user in a certain atmosphere.



\bigskip
\noindent
\textbf{Funding}  No funding was received for conducting this study.

\noindent
\textbf{Data availability}  
Data reported in this article, including both generated images and videos, are available at
https://github.com/LeomPina/AI-Driven-Music-Visual-Generation-Using-Diffusion-Models.
Other datasets generated during the study are available from the corresponding author on reasonable request. 

\section*{Declarations}
\textbf{Competing interests}  The authors have no financial or non-financial competing interests to declare relevant to the content of this article.

\bibliography{sn-bibliography}
\end{document}